\documentclass[reprint,amsmath,amssymb,aps]{revtex4-1}

\usepackage{graphicx}
\usepackage{dcolumn}
\usepackage{bm}
\usepackage{gensymb}

\begin{document}

\preprint{APS/123-QED}

\title{Dynamic Sealing Using Magneto-Rheological Fluids}

\author{Youzhi Liang}
	\email{Corresponding author\\YOUZHIL@mit.edu}
	\affiliation{Department of Mechanical Engineering, Massachusetts Institute of Technology, 77 Massachusetts Avenue, Cambridge, Massachusetts 02139, USA\\}
\author{Jose R. Alvarado}
	\affiliation{Department of Mechanical Engineering, Massachusetts Institute of Technology, 77 Massachusetts Avenue, Cambridge, Massachusetts 02139, USA\\}
\author{Karl D. Iagnemma}
	\affiliation{Department of Mechanical Engineering, Massachusetts Institute of Technology, 77 Massachusetts Avenue, Cambridge, Massachusetts 02139, USA\\}
\author{A. E. Hosoi}
	\affiliation{Department of Mechanical Engineering, Massachusetts Institute of Technology, 77 Massachusetts Avenue, Cambridge, Massachusetts 02139, USA\\}

\date{\today}

\begin{abstract}
Micropumps are microfluidic components which are widely used in applications such as chemical analysis, biological sensing and micro-robots. However, one obstacle in developing micropumps is the extremely low efficiency relative to their macro-scale counterparts. This paper presents a dynamic sealing method for external gear pumps to reduce the volumetric losses through the clearance between the tips of gears and the housing by using magneto-rheological (MR) fluids. By mitigating these losses, we are able to achieve high efficiency and high volumetric accuracy with current mechanical architectures and manufacturing tolerances. Static and dynamic sealing using MR fluids are investigated theoretically and experimentally. Two Mason numbers $Mn\left(p\right)$ and $Mn\left(\Omega\right)$ which are defined in terms of pressure gradient of the flow and velocity of the moving boundary respectively are used to characterize and evaluate the sealing performance. A range of magnetic field intensities is explored to determine optimal sealing effectiveness, where effectiveness is evaluated using the ratio of volumetric loss and friction factor. Finally, we quantify the effectiveness of this dynamic sealing method under different working conditions for gear pumps. 
\end{abstract}
\maketitle

\section{\label{sec:level1}Introduction\protect\\}
Micropumps are miniaturized pumping devices that are usually manufactured by MEMS micromachining technologies \cite{abhari2012comprehensive,tay2002microfluidics}. In recent years, the target applications have expanded owing to the integration of novel physical principles and the invention of new fabrication methods. Micropumps are commonly used in chemical analyses, biological sensing, drug delivery and micro-robots \cite{junwu2005design,cui2007study,lintel1988piezoelectric}.

\begin{figure}[bh!]
\includegraphics[scale=0.33]{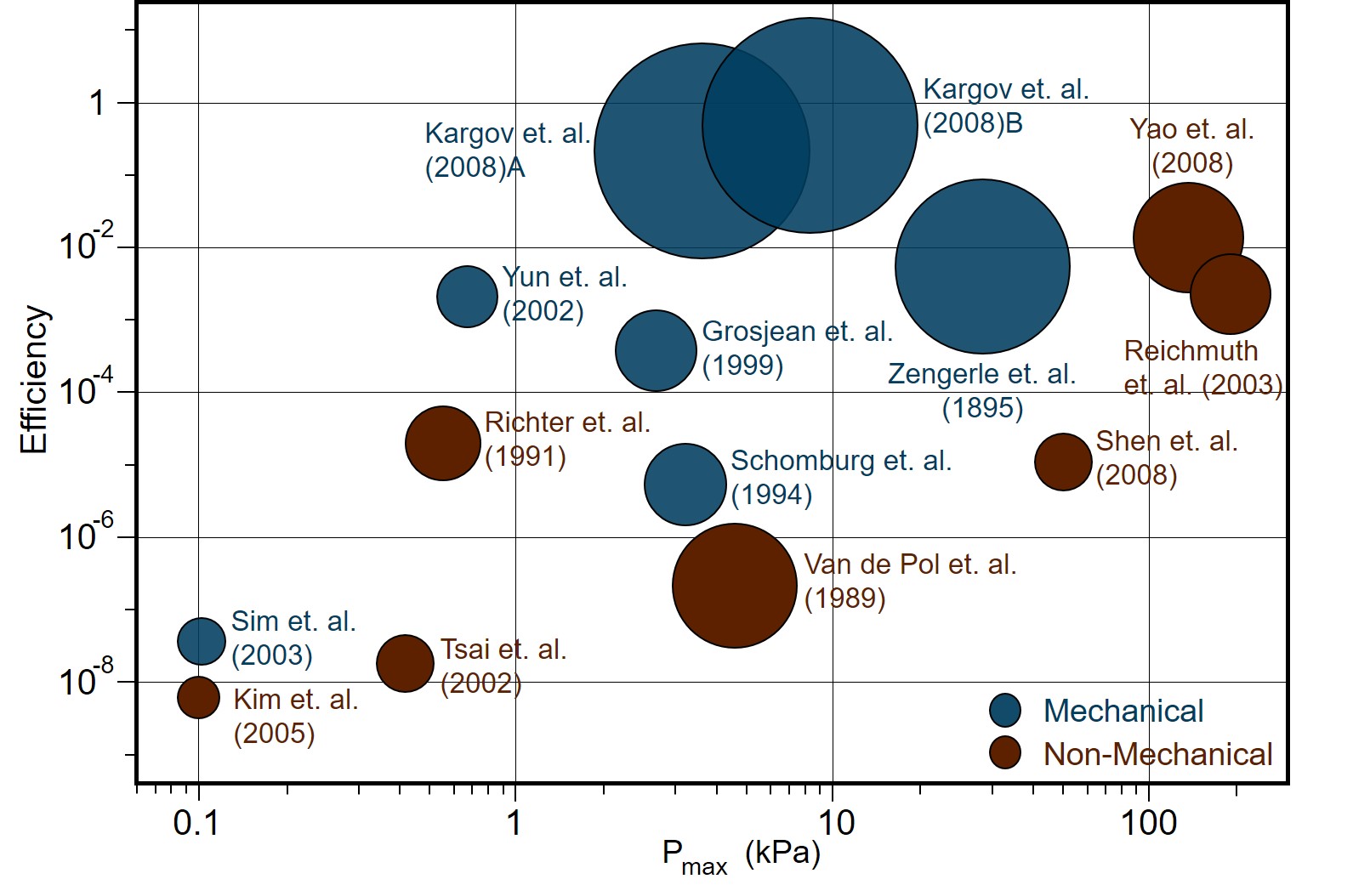}
\caption{\label{fig:efficiency}Review of efficiency versus maximum pressure for existing small-scale pumping strategies. The size of the symbol depicts the characteristic length scale of the pump package; the location of the center depicts the effeciency versus the pressure. For reference, Sim {\em et.~al} is 72 mm$^3$ and Kargov {\em et.~al} is 18.5 cm$^3$. Pumps shown here include: Sim {\em et.~al}: Micropump with flap valves \cite{sim2003phase}, Kim {\em et.~al}: Electromagnetic pump \cite{kim2005fabrication}, Yun {\em et.~al}: Surface-tension driven pump \cite{yun2002surface}, Richter {\em et.~al}: Electrohydrodynamic pump \cite{richter1991micromachined}, Tsai {\em et.~al}: Thermal-bubble-actuated pump \cite{tsai2002thermal}, Kargov {\em et.~al}: Gear pump \cite{kargov2008development}, Grosjean {\em et.~al}: Thermopneumatic pump \cite{yang1999design}, Schomburg {\em et.~al}: Pneumatic chamber pump \cite{rapp1994liga}, Van de Pol {\em et.~al}: Thermopneumatic pump \cite{van1990thermopneumatic}, Zengerle {\em et.~al}: Electrostatic pump \cite{zengerle1995bidirectional}, Shen {\em et.~al}: Electromagnetic pump \cite{shen2008high}, Yao {\em et.~al}: Electroosmotic pump \cite{yao2003porous}, Reichmuth {\em et.~al}: electrokinetic pump \cite{reichmuth2003increasing}.}
\end{figure}

Unfortunately, miniaturization comes at a cost and nearly all micropumps suffer from low efficiency. The reported efficiencies of the available micropump technologies are shown in Fig.~\ref{fig:efficiency}. Typically, the overall efficiency of a micropump is determined by a combination of four components: volumetric efficiency, hydraulic efficiency, mechanical efficiency and electrical efficiency. Out of these four, volumetric losses and hydraulic losses dominate at small scales. As the size of the system decreases, the volumetric efficiency decreases since the same dimensional and geometric tolerances result in a larger fractional loss. Furthermore, in terms of hydraulic efficiency, the Reynolds number decreases as the systems size decreases, resulting in larger viscous losses.

For external gear pumps, the volumetric losses are roughly proportional to the pressure gradient assuming a quasi-steady fully developed low Reynolds number flow across the clearance between the housing and the gear tips \cite{dopper1997micro}. Thus, the effeciency may be extremely low when the pump is operating under high pressure gradient conditions. The volumetric leakage between the tips of the gears and across the side plates is typically considered to comprise the largest proportion of the total efficiency loss in external gear pumps \cite{totten2011handbook,merritt1967hydraulic}. Various end wear plates have been studied and designed to reduce the leakage across the side plates \cite{hooke1984end}. However, studies that consider volumetric losses between the tip of the gear teeth and the housing are relatively rare. Sealing is even more challenging for micro-scale gear pumps due to the limits of manufacturing precision. With precise manufacturing techniques and tight tolerances, the volumetric loss could be reduced. But in that case the mechanical friction between the housing and the gears will increase and small clearances may also make the pump more vulnerable to vibrations. Therefore, we propose to develop a dynamic sealing method using magneto-rheological (MR) fluids that can operate with the current mechanical architectures and manufacturing tolerances.

Magnetorheological (MR) fluids are materials that exhibit a reversible change in rheological properties with the application of an external magnetic field, which can result in a rich range of physical properties \cite{furst1998particle, tao1998structural, zhu1996role, jolly1999properties}. In engineering applications, they were initially used by Jacob Rabinow in the design of a clutch in the late 1940s \cite{goncalves2006review}. In more recent years, MR fluids have found further applications and commercial success \cite{ginder1996rheology}. Typical operational modes for MR fluid application are the pressure driven flow mode and the direct shear mode \cite{tao1998structural,jolly1999properties}. The most common application is a mechanical damper, which yields appealing features such as low-power consumption, force controllability and rapid response \cite{lee2000control, kwok2007bouc}. In particular, automotive dampers with these properties have been widely investigated \cite{lee2000control,kim1999vibration,sassi2005innovative}. The other common use of MR fluids is the development of MR valves. In addition, high efficiency, miniaturized MR valves have been achieved \cite{yoo2002design,guo2003finite}. 

\begin{figure}[b]
\includegraphics[scale=0.65]{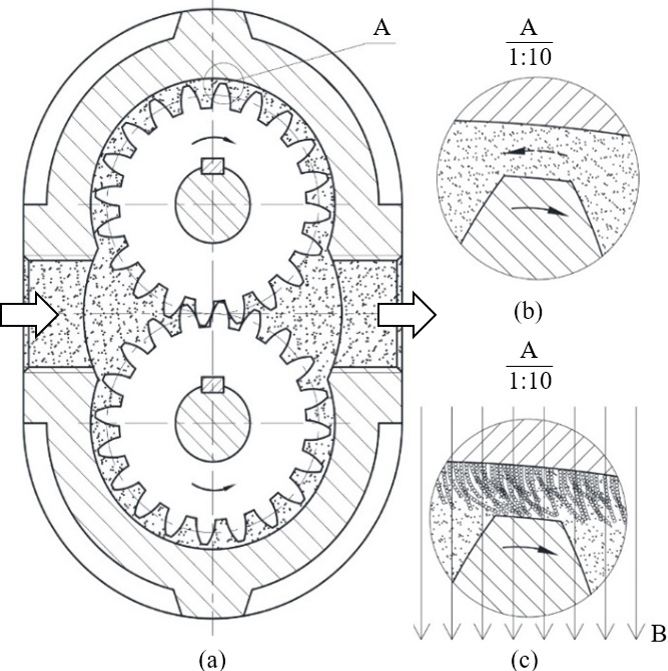}
\caption{\label{fig:gearpump}(a) Schematic of an external gear pump; (b) Volumetric loss caused by back-flow in-between the gear tooth and the housing (indicated by dashed arrow); (c) Proposed dynamic sealing method using MR fluids in presence of magnetic field $B$.}
\end{figure}

A schematic of a typical external gear pump is shown in Fig.~\ref{fig:gearpump}. The pressure of the outlet is larger than that of inlet, resulting in back-flow aross the gap between the tips of the gears and the pump housing, as shown in Fig.~\ref{fig:gearpump} (b). Meanwhile in Fig.~\ref{fig:gearpump} (c), subjecting MR fluid to an external magnetic field causes magnetic-induced dipoles to aggregate in the vicinity of the housing, which prevents the back-flow. This design have the potential to dispense of precise manufacturing and solve the challenge of controlling the clearance between the housing and gear tips.

Previous research primarily focused on MR fluids in either Couette flow or Poiseuille flow, usually within the scope of high shear stress which arises from either a large pressure differential or large exerted force \cite{jolly1999properties,goncalves2006review,ginder1996rheology,tao1998structural,jolly1999properties,kwok2007bouc,lee2000control,kim1999vibration,sassi2005innovative,yokota1999pressure,yoo2002design,guo2003finite}. By contrast, much less is known about the physics of MR fluids subject to the combination of Couette and Poiseuille flow. In this study, we investigate the performance of dynamic seals of MR fluid chains subject to shear-driven flows from gear motion and simultaneously to pressure-driven flows from back-flow. We compare experimental results to a model which incorporates two dimensionless Mason numbers, one from Couette flow and one from Poiseuille flow.

\section{\label{sec:level1}Method\protect\\}
\subsection{\label{sec:level2}Experiments for MR Fluid in Poiseuille Flow or Couette Flow}
\begin{figure}
\includegraphics[scale=0.75]{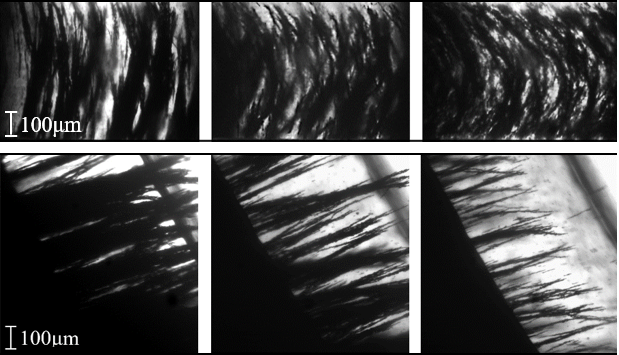}
\caption{\label{fig:poiseuille} (Top) Deformation of chains of induced dipoles in Poiseuille flow. (Bottom) Deformation of chains of induced dipoles in Couette flow. (The contrast of the pictures has been enhanced.) (Top) Flow rates from the left to the right: 0.1 ml/min, 1 ml/min, 10 ml/min. Channel width: 0.7 mm. (Bottom) Rotational speed from the left to the right: 0 rps, 0.4 rps, 0.8 rps. Radius of the disk: 25 mm, Channel width: 0.7 mm. }
\end{figure}

In order to visualize the effect of Poiseuille flow on the morphology of the MR chains, we designed a experimental system specialized for visualization. We built a micro-channel network made of a silicon slide, which is laser-cut and sandwiched between two transparent acrylic plates. MR fluid is from Lord. The volume fraction of MR particles is diluted to be 1\%. The flow was driven by a pressure gradient using a syringe pump to control the flow rate. Typical deformations in the channel for different flow rates are displayed in Fig.~\ref{fig:poiseuille} (Top). The images suggest that the deformation of the magnetic chains increases as flow rate increases until the magnetic chains finally collapse. Note that with low flow rate, magnetic chains tend to aggregate in bunches with very little deformation. These chains appear to attach in the vicinity of the walls of the channel. As the flow rate increases, the chains are more clearly deformed and segregated.

To observe the deformation of MR chains under Couette flow, we built another experimental system. We run the experiments without an adverse pressure gradient. Results are shown in Fig.~\ref{fig:poiseuille} (Bottom); the left black area of each figure depicts a roughened stationary surface, and the black line on the upper right depicts the surface of the disk. As the rotational speed increases, the density of magnetic brushes decreases with a larger curvature.

\subsection{\label{sec:level2}Experiments for MR Fluid in Poiseuille-Couette Flow}
\begin{figure}
\includegraphics[scale=0.45]{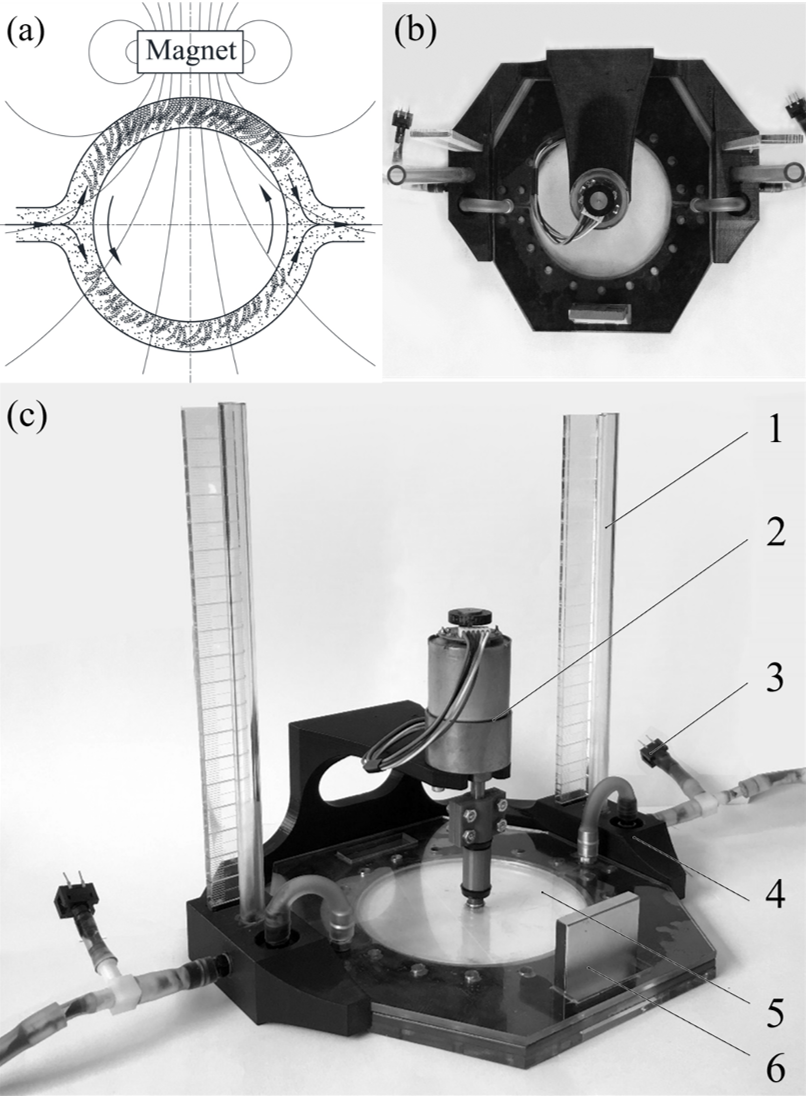}
\caption{\label{fig:setup}(a) Schematic of the experiment designed to study the dynamic sealing performance using MR fluid; (b) Top view of the experimental setup. (c) Perspective view of the experimental setup. 1: pitot tube, 2: motor, 3: pressure sensor, 4: frame, 5: disk (inside), 6: magnet.}
\end{figure}

To model the interaction between the gear tooth and the housing, we built a simplified experimental system, as shown in Fig.~\ref{fig:setup}. Panel (a) depicts a schematic of the underlying design, where fluid enters the inlet on the left, then bifurcates into two slots. The slot is used to mimic the clearance between the tip of the gears and the housing. A rotating disk is utilized to mimic one gear tooth.

Fig.~\ref{fig:setup} (b) and (c) show snapshots of the experimental model system, which mainly consists of frame, motor, disk, pitot tube, magnet and pressure sensor. The frame, which designed to secure other components, contains a cavity, which connects the tube fitting, the pitot tube and the tube connected to the slots in the middle. A laser cut acrylic disk, driven by the motor, is sandwiched between two transparent plates with slots to locate the magnet. The carrier fluid of the MR fluids is silicone oil (Gelest, 100 cSt). MR fluid is from Lord. The volume fraction of MR particles is diluted to be 10\%. The magnets have a surface field of 1895 Gauss (NdFeB, Grade N42, 2.44 oz.).

We used a variable voltage power supply to power both the sensors and the motor (Pololu 12 V), using voltages of 10.5 V and from 0 V to 40 V respectively. Pressure data acquired from the sensors were sent to Labview via National Instruments I/O. Motor speed was acquired from the encoder of the motor and sent to the Arduino built-in serial monitor via Arduino Uno.

\section{\label{sec:level1}Results and Discussion\protect\\}
\subsection{\label{sec:level2}Model for MR Fluid in Poiseuille-Couette Flow}
\begin{figure}
\includegraphics[scale=0.35]{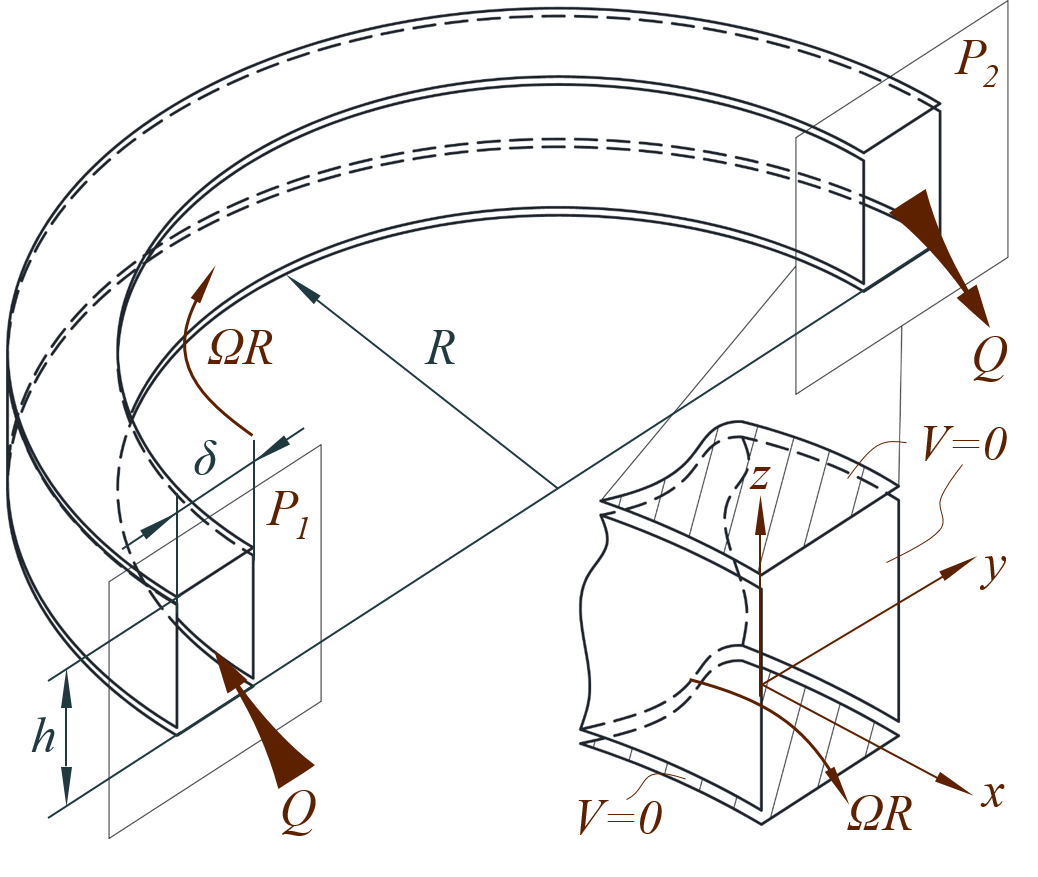}
\caption{\label{fig:annuli}Schematic of one slot. In the experiment, the radius of the inner annuli is $R = 50$ mm, the width of the slot is $\delta = 2$ mm, the height is $h = 4$ mm. The inner annulus rotates at angular velocity of $\Omega R$; the outer annulus is stationary. $P_1$ and $P_2$ denote different pressures.}
\end{figure}

The experimental setup, shown in Fig.~\ref{fig:setup}, can be modeled as two slots in parallel in the presence of different magnetic field intensity. A schematic of one slot is shown in Fig.~\ref{fig:annuli}. This half can be simplified as a straight channel with the reference frame attached shown in the partial enlarged view in Fig.~\ref{fig:annuli}, based on the fact that the aspect ratio $\delta/R = 2$ mm$/$50 mm $\ll$ $1$. Thus, we consider two straight slots in parallel. The theoretical results are calculated numerically to account for the square channel cross-section \cite{rowe1970measurements}. The Reynolds number $Re_{\delta} = \frac{\rho U\delta}{\mu}=0.05 \ll 1$, so the inertia of the MR fluid is negligible.
\begin{figure}
\includegraphics[scale=0.75]{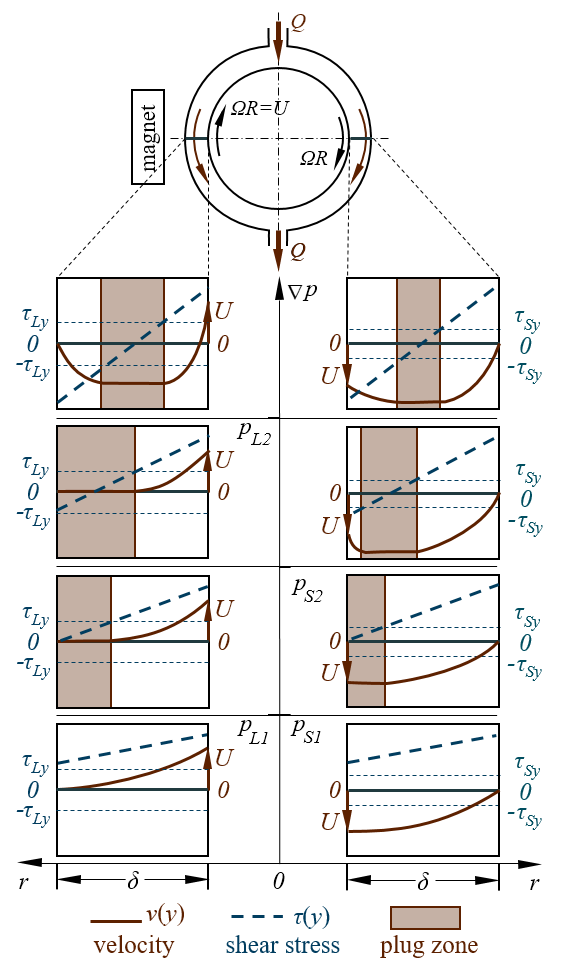}
\caption{\label{fig:velocityprofile}Schematic of the profiles for velocity, shear stress and plug zone of Poiseuille-Couette flow of MR fluid in both slots, exposed in larger magnetic field intensity (left) and smaller magnetic field intensity (right) respectively, as pressure gradient $\nabla P$ increases. $\delta$ indicates the width of the channel; the arrow in the profile plots indicates the velocity $U$ of the moving wall. $p_{L1}$ and $p_{L2}$ represents the transition pressure from one-region mode to two-region mode and from two-region mode to three-region mode respectively, in presence of larger magnetic field intensity; $p_{S1}$ and $p_{S2}$ represents those in presence of smaller magnetic field intensity.}
\end{figure}

The flow is driven by both a pressure gradient and a moving wall. In the limit of low Reynolds number, the conservation of momentum equation for steady, laminar flow, in the $x$-direction, reduces to:
$$\frac{dp}{dx}=\frac{d\tau_{yx}}{dy} ,$$
where $p$ is the mechanical pressure, $\tau_{yx}$ is the shear stress.

The MR fluid is modeled as a Bingham fluid in this paper. Due to the distribution of magnetic field intensity, the yield stress is larger in the slot closer to the magnet than that in the further one. The constitutive relationship can be expressed as:
\begin{eqnarray*}
\tau_{yx}=\left(\mu+\frac{\tau_y}{|\dot{\gamma}|}\right)\dot{\gamma} &;& |\tau| > \tau_y\\
\dot{\gamma}=0 &;& |\tau| \leq  \tau_y ,
\end{eqnarray*}
where $\mu$ is the viscosity of the MR fluid, $\tau_y$ is the yield stress, and $\dot{\gamma}$ is the shear rate.\\
We furthermore have the following boundary conditions on the inner and outer walls of the channel:
\begin{eqnarray*}
v_x|_{y=0}&=&U\\
v_x|_{y=\delta}&=&0 ,
\end{eqnarray*}
where $v_x$ is the velocity of the fluid in $x$-direction and $U$ is the velocity of the inner wall.

To characterize the behavior of the dipole chains, we use the Mason number, which has been commonly considered in prior studies \cite{melle2003microstructure,du2016modified,van2014microfluidics,becnel2014mason}. In our study, we define two Mason numbers: one which is the ratio between the shear forces and the magnetic interaction forces in Poiseuille flow, and another one for Couette flow. The magnetic interaction forces are characterized by the yield stress $\tau_y$ \cite{becnel2015nondimensional,sherman2015relating}.
\begin{eqnarray*}
Mn\left(p\right) &=& \frac{\delta}{\tau_y}\left(-\frac{dp}{dx}\right)\\
Mn\left(\Omega\right) &=& \frac{\tau_y \delta}{\mu R \Omega} .
\end{eqnarray*}
To non-dimensionlize the governing equation, the other dimensionless variables are defined as follows:
$$y^* = \frac{y}{\delta}; \tau^* = \frac{\tau_{yx}}{\tau_y}; v^* = \frac{v_x}{R \Omega}; U^* = \frac{R \Omega}{| R \Omega |} .$$
Substituting the dimensionless variables into the conservation of momentum equation and the constitutive equation yields:
\begin{eqnarray*}
\frac{d\tau^*}{dy^*} + Mn\left(p\right) = 0 &;&\\
\tau^* = \frac{1}{Mn \left( \Omega \right)} \frac{dv^*}{dy^*} + sgn \left( \frac{dv^*}{dy^*} \right)&;& |\tau^* | > 1\\
\frac{dv^*}{dy^*} = 0 &;& |\tau^* | \leq 1 .
\end{eqnarray*}
The boundary conditions become:
	$$v^*|_{y^*=0}=U^*;v^*|_{y^*=1}=0 .$$
	
The velocity profiles can be computed from the governing equation and the associated boundary conditions, and can be categorized into three modes: (i) a one-region mode, (ii) a two-region mode and (iii) a three-region mode \cite{gjerstad2012simplified,liu2010axial}. (i) The one-region mode occurs when the pressure gradient is small and the velocity of the boundary is relatively large. The fluid stress is larger than the yield stress of the Bingham fluid across the entire slot, so MR chains cannot form. The velocity profile in the one-region mode is identical to that of a Newtonian fluid in Poiseuille-Couette flow. (ii) The two-region mode occurs as the pressure gradient increases, which increases the slope of the stress distribution. In the region where the fluid stress is smaller than the yield stress, a plug zone will occur, where MR chains form and the velocity profile resembles plug flow. In two-region mode, the plug zone is anchored to the surface nearest the magnet, whereas in the region at the opposing surface the MR particles are prevented from aggregating, similarly to one-region mode. (iii) Finally, the three-region mode occurs as the pressure gradient increases even further. Under such conditions, the plug zone will detach from the wall and move to the middle of the channel, surrounded by Newtonian regions on either side. Considering these three types of modes for the two slots in our experimental study, there are four possible combinations of velocity profiles in this study, as shown in Fig.~\ref{fig:velocityprofile}. The slot closest to the magnet is in the presence of a higher magnetic field, resulting in a larger yield stress $\tau_{Ly}$. In our study, $\tau_{Ly}$ is about four times larger than $\tau_{Sy}$.\\

The average velocity of the fluid in the one-region mode is given by:
	$${\overline{v}}^* = \displaystyle{\frac{1}{12}}  Mn \left(\Omega\right) Mn \left(p\right) + \frac{1}{2}U^* .$$
The average velocity of the fluid in the two-region mode depends on the sign of $Mn(p)$ and $Mn(\Omega)$. When $Mn(p)$ has the same sign as $M(\Omega)$, we have:
	$${\overline{v}}^* = U^* - \frac{{U}^*}{3} \sqrt{\frac{{2U}^*}{Mn\left(\Omega\right) Mn\left(p\right)}} .$$
and when $Mn(p)$ has the opposite sign to $Mn(\Omega)$, we have:
	$${\overline{v}}^* = \frac{{U}^*}{3} \sqrt{\frac{{-2U}^*}{Mn\left(\Omega\right) Mn\left(p\right)}} .$$
The average velocity of the fluid in the three-region mode is given by:
\begin{eqnarray*}
{\overline{v}}^* = &\displaystyle{\frac{1}{12}}& Mn\left(\Omega\right) Mn\left( p \right) \left( 1 - \frac{3}{|Mn\left(p\right)|} + \frac{4}{|Mn\left(p\right)|^3}\right) \\
&+& \frac{U^*}{2} \pm \frac{1}{Mn(\Omega)(2 \mp Mn(p))^2} .
\end{eqnarray*}
The transition pressure from one-region mode to two-region mode and from two-region mode to three-region mode can also be computed and are found to be quantities $Mn(p)_{R1}$ and $Mn(p)_{R2}$ respectively:
\begin{eqnarray}
{Mn(p)}_{R1} &=& \frac{2}{Mn(\Omega)} \\
\label{eq:one}
{Mn(p)}_{R2} &=& 2+\frac{1}{Mn(\Omega)} + \sqrt{\frac{1}{{Mn(\Omega)}^2} + \frac{4}{Mn(\Omega)}} .
\label{eq:two}
\end{eqnarray}			

\subsection{\label{sec:level2}Experimental Results for MR Fluid in Poiseuille-Couette Flow}
\begin{figure}
\includegraphics[scale=0.37]{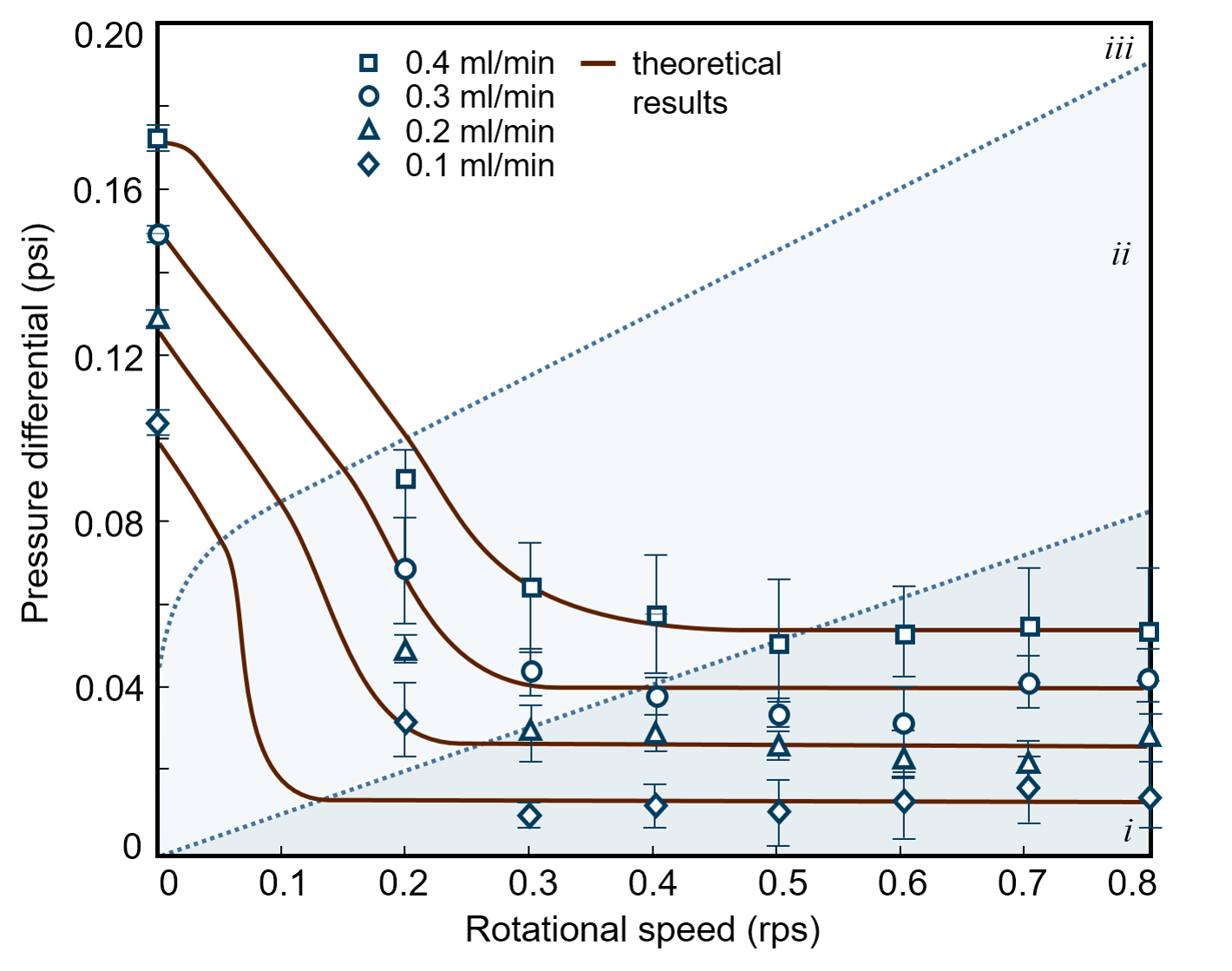}
\caption{\label{fig:dataanalysis}Experimental and theoretical results. Pressure differential (psi) as a function of rotational speed (rps), for flow rates from 0.1 ml/min to 0.8 ml/min. Each point corresponds to the mean from three iterations of experiments, with error bars indicating standard deviation. \textit{i}: one-region mode; \textit{ii}: two-region mode; \textit{iii}: three-region mode.}
\end{figure}

\begin{figure}
\includegraphics[scale=0.38]{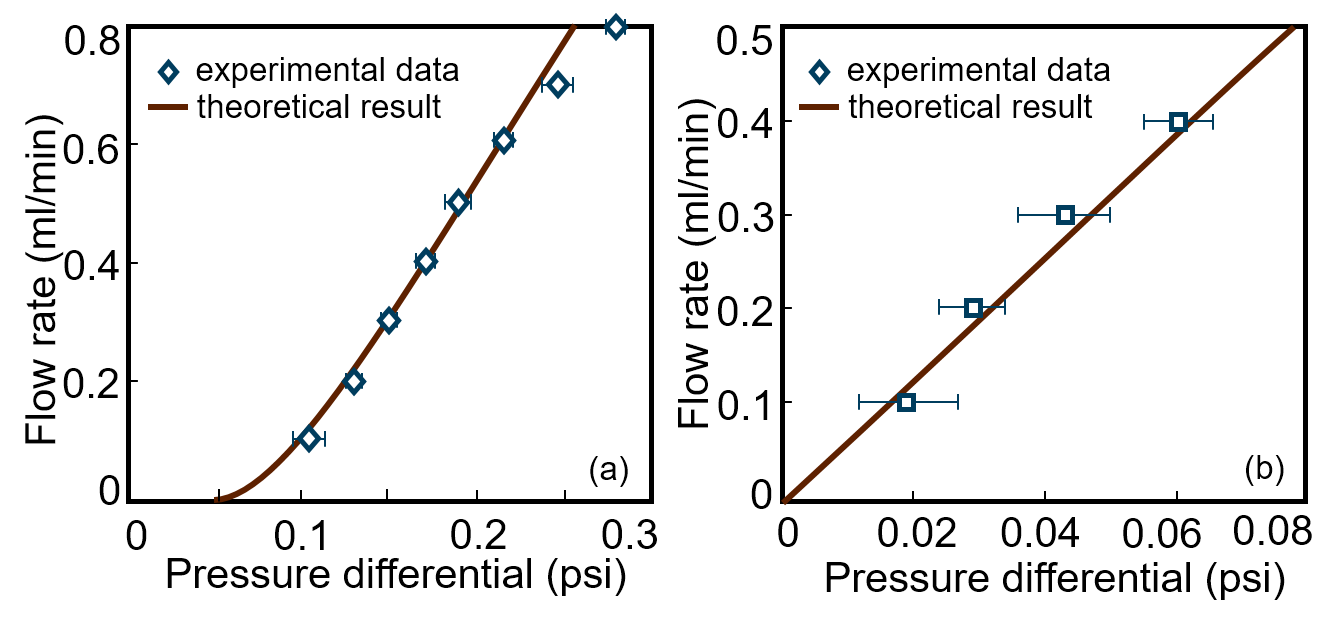}
\caption{\label{fig:limit}Experimental data and theoretical results of Poiseuille flow of MR fluids (a) as Bingham fluid when rotational speed of the disk is zero, and (b) in one-region mode when the rotational speed is sufficiently large. Flow rates (ml/min) as a function of pressure differential (psi).}
\end{figure}

We investigate the performance of the dynamic seals using the experimental setup shown in Fig.~\ref{fig:setup} (b). The flow rate is controlled by a syringe pump, ranging from 0.1 ml/min to 0.8 ml/min. For each given flow rate, the rotational speed of the motor is varied from 0 rps to 0.8 rps. The pressure differential is given by two pressure sensors located at the inlet and outlet. The results are shown in Fig.~\ref{fig:dataanalysis} (a). As the rotational speed increases, the pressure differential decreases abruptly from the static state to the dynamic state, and settles to a steady state.

We apply numerical methods to calculate the velocity profile in the rectangular cross-section, and integrate the velocity profile in the cross-section to get the total flow rate. We first consider Couette flow as a simple example. In the case of two parallel infinite plates, the velocity decreases linearly away from the moving wall. In the real experimental setup, as shown in Fig.~\ref{fig:setup} and Fig.~\ref{fig:annuli}, the aspect ratio $h:\delta=4$ mm$:2$ mm$=2:1$ with a clearance 0.127 mm due to the gasket for sealing. The ratio of the average velocity of the real case to that of the Couette case is $0.241:0.5=0.482$. Pressure losses between the sensors and the inlet and outlet have been taken into account.  In our experiments, a significant pressure loss occurs due to the Poiseuille flow between the sensors and the inlet and outlet, four 90$\degree$-elbows and the cavities which connects the pitot tube and the two tubings. The distance between the pressure sensor and the outlet is $50$ mm; the inner diameter of the tube is $d=3$ mm. The Reynolds number $Re_d = \frac{\rho Ud}{\mu}\sim 10^{-3} \ll 1$, so inertial effects are negligible. For elbows, the pressure loss is estimated from empirical equation.  For 90$\degree$-elbow curved, the equivalent length is $L_{eq}=16d$ \cite{menon2004piping}.

Our mathematical Bingham model of MR fluid agrees well with the experimental results, as shown in Fig.~\ref{fig:limit}. One limiting condition is when the rotational speed is zero, which corresponds to the classic Poseuille flow for Bingham fluid. In our experiment, this condition can be treated as two slots for Bingham Poiseuille flow in parallel with different yield stresses (Fig.~\ref{fig:limit} (a)). The other limiting condition is that the rotational speed of the disk is fast enough so that the flow in both slots are in the one-region mode. Thus, the velocity profile of MR fluid is identical as that of Poiseuille-Couette flow of a Newtonian fluid. Because the directions of the Couette flow are opposite in the parallel slots, the flow rate induced by Couette flow is actually canceled out. Thus, the flow rate as a function of pressure gradient is linear, as shown in Fig.~\ref{fig:limit} (b).

\begin{figure}
\includegraphics[scale=0.7]{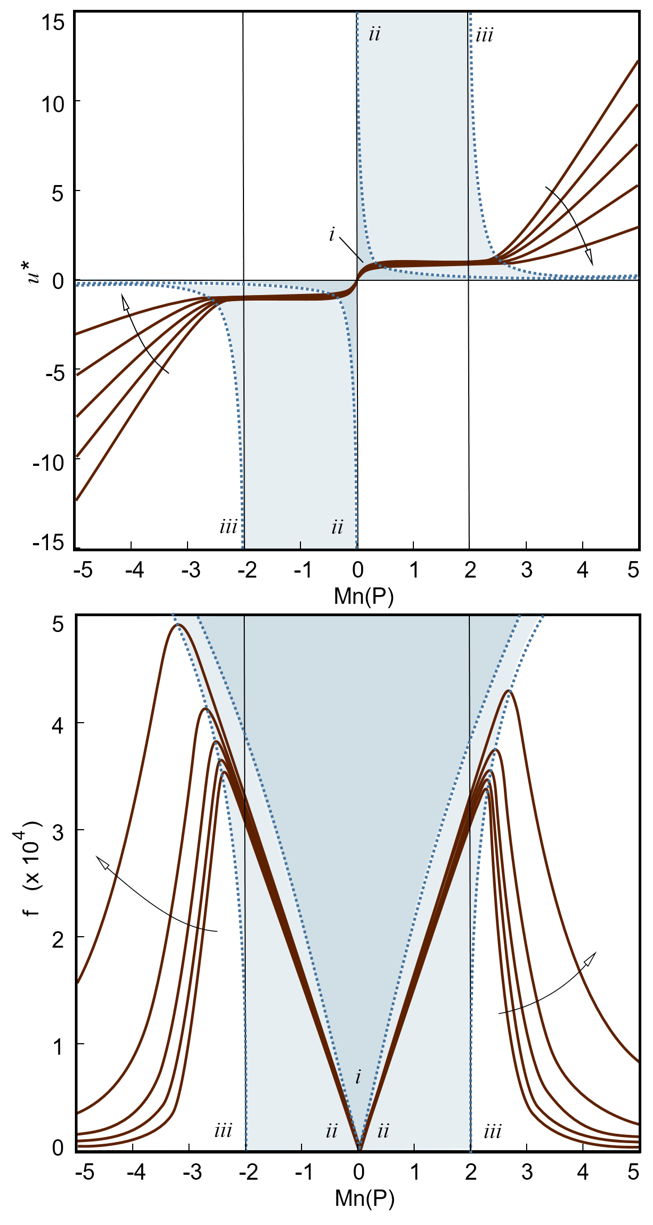}
\caption{\label{fig:evaluation}(a) Ratio of volumetric loss to the nominal flow rate of a gear pump as a function of $Mn(p)$, for $Mn(\Omega)$ equal 0.5, 1, 1.5, 2, 2.5. The arrow indicates the direction $Mn(\Omega)$ increases. (b) Friction factor as a function of $Mn(p)$, for $Mn(\Omega)$ equal 0.5, 1, 1.5, 2, 2.5. The arrow indicates the direction $Mn(\Omega)$ increases. \textit{i}: one-region mode; \textit{ii}: two-region mode; \textit{iii}: three-region mode. The dot lines indicate the transition for the velocity profile to transit from one mode to another. }
\end{figure}
\subsection{\label{sec:level2}Optimal Magnetic Field Intensity}
In terms of the design and application for external gear pumps, we consider two performance metrics which evaluate the performance of dynamic seals. The first performance metric is given by the ratio of volumetric flow rate loss to the nominal volumetric flow rate of the gear pump. The nominal volumetric flow rate is proportional to the angular speed of the gear. Therefore, the dimensionless group $u^* = \displaystyle{\frac{v}{R\Omega}}$ can be used to characterize the sealing effectiveness of MR fluid, where $v$ is the average velocity of the back-flow rate in the clearance of the gear pump, $R\Omega$ is proportional to the volumetric flow rate pumped by the gear pump. As shown in Fig.~\ref{fig:evaluation} (a), to achieve higher effectiveness, $u^*$ should be designed to be as small as possible. $Mn(p)_{R1}$ is the transition point of the velocity profile from one-region mode to two-region mode for both slots, because $Mn(p)_{RS1}$ equals $Mn(p)_{RL1}$. $Mn(p)_{SR2}$, $Mn(p)_{LR2}$ are the transition points of the velocity profile from two-region mode to three-region mode for the slots in the presence of larger and smaller magnetic field intensity respectively. We find that when $Mn(p)$ is larger than $Mn(p)_{SR2}$, $u^*$ dramatically increases. Thus, to ensure a small volumetric loss, $Mn(p)$ should be smaller than $Mn(p)_{SR2}$.

The second performance metric comes from the energy loss in both of the slots, which can be characterized by the friction factor $f=\displaystyle{\frac{p}{\frac{1}{2} \rho v^2}}$. To achieve the optimal sealing performance, the friction factor needs to be maximized, indicating that the back-flow between the gear teeth and the housing will experience as much energy loss as possible. As shown in Fig.~\ref{fig:evaluation} (b),  the maximum friction factor can be achieved around $Mn(p)_{SR2}$, which is the $Mn(p)$ of the transition point from two-region mode to three-region mode for the slot in presence of the smaller magnetic field intensity.

\begin{figure}
\includegraphics[scale=0.55]{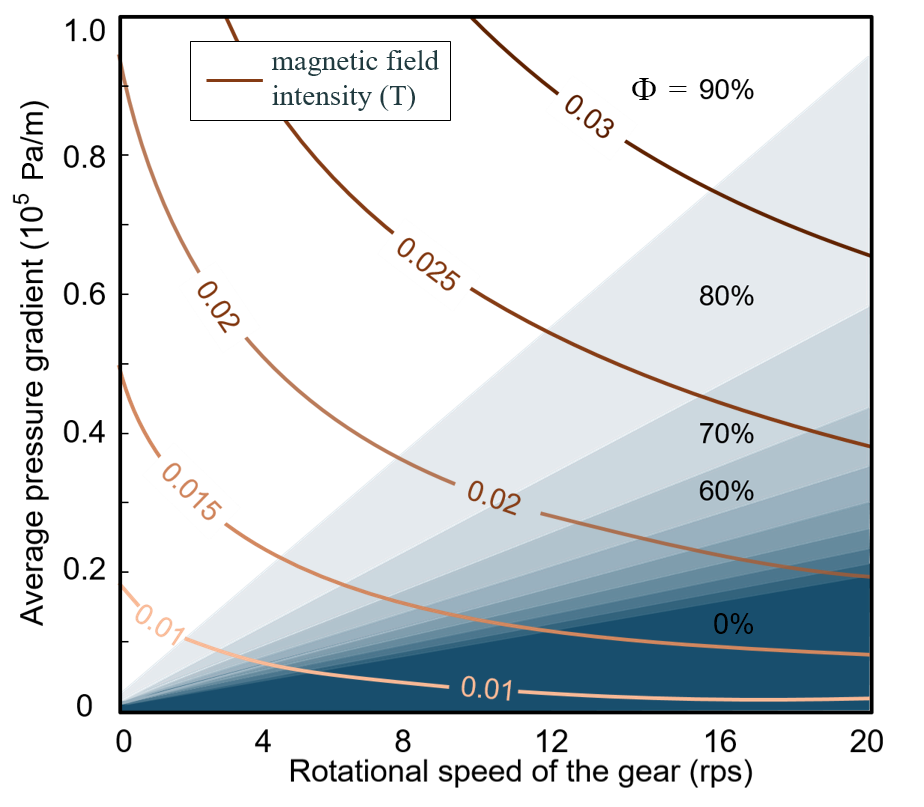}
\caption{\label{fig:conclusion}Optimal magnetic field intensity (T). Solid line indicates the magnetic field intensity distribution: from 0.01 T to 0.03 T at given specific condition. Shaded area indicates the percentage of the reduction in volumetric loss from $\Phi=$ 90\% (white) to 0\% (black).}
\end{figure}

Upon considering the two performance metrics mentioned above, the optimal sealing performance can be achieved at the transition of two-region mode to three-region mode. Thus, at any given nominal work condition of external gear pump, the magnetic field intensity can be tuned to make the yield stress satisfy Eq.~(\ref{eq:two}), which can be expressed explicitly by the following equation:
$$\tau_y=\frac{1}{2}\frac{{\left(\displaystyle{\frac{dp}{dx}}\delta\right)}^2-2\mu\displaystyle{\frac{dp}{dx}}R\Omega}{\displaystyle{\frac{dp}{dx}}\delta-\sqrt{2\mu\displaystyle{\frac{dp}{dx}}R\Omega}}.$$

The relationship between magnetic field intensity ($B$) and yield stress ($Pa$) of MR fluid has been studied in prior studies: \cite{jolly1999properties}
$$\lg{B}=\frac{4}{7}\lg{\tau_y}+\frac{4}{7}\lg{\left(9\times10^{-4}\right)} ,$$
where $B$ is the magnetic field intensity, $\tau_y$ is the yield stress.

We define a ratio $\Phi$ as a metric for the effectiveness of dynamic seals using MR fluid:
$$\Phi = \frac{Q_{Oil}-Q_{MR}}{Q_{Oil}} ,$$
where $Q_{Oil}$ is the volumetric loss using general pump oil, $Q_{MR}$ is the volumetric loss using MR fluid with the same viscosity as the general pump oil.

The optimal magnetic field intensity is shown in Fig.~\ref{fig:conclusion}. It suggests that dynamic sealing using rheological fluid will achieve the optimal sealing effectiveness under a high pressure gradient and relatively low rotational speed, which would reduce the volumetric loss to over 90\%.

\section{\label{sec:level1}conclusion\protect\\}
Volumetric loss accounts for the extremely low efficiency of small-scale gear pumps. In order to reduce the volumetric loss without introducing larger friction, tighter manufacturing tolerances, or vulnerability to vibrations, we introduced a method where magnetorheological fluid is activated in the vicinity of the clearance between gear and housing to create a dynamic seal.

We verified the Bingham fluid model for MR fluids, and have accounted for the combined Poiseuille-Couette flow at low Reynolds number in the application of sealing in external gear pumps. We furthermore found four possible combinations of the velocity profiles given by two modified Mason numbers $Mn\left(p\right)$ and $Mn\left(\Omega\right)$. 

We determined the dependence of optimal magnetic field intensity on the pressure gradient and rotational speed of the gear. The optimal magnetic field intensity corresponds to the transition for the velocity profile of MR fluid to transit from three-region mode to two-region mode. Our dynamic sealing method using MR fluid reduces volumetric loss above 90\% when the pressure gradient is large; that is, when the hydraulic actuation system is under heavy load at low speed. Besides application for reducing the volumetric loss in the clearance, our method can also be applied for reducing the loss between the housing and the sides of gears for all types of gear pumps.

\begin{acknowledgments}
A.E.H. acknowledges support from the Defense Advanced Research Projects Agency under grant number DARPA W31P4Q-13-1-0013. We gratefully thank Marie Baumier for insightful discussions.
\end{acknowledgments}

\bibliography{DynamicSealingUsingMagneto-RheologicalFluid}

\providecommand{\noopsort}[1]{}\providecommand{\singleletter}[1]{#1}%
\begin{thebibliography}{10}

\bibitem{abhari2012comprehensive}
Farideh Abhari, Haslina Jaafar, and Nurul Amziah~Md Yunus.
\newblock A comprehensive study of micropumps technologies.
\newblock {\em Int. J. Electrochem. Sci}, 7:9765--9780, 2012.

\bibitem{tay2002microfluidics}
Francis~EH Tay.
\newblock {\em Microfluidics and BioMEMS applications}.
\newblock Springer, 2002.

\bibitem{junwu2005design}
Kan Junwu, Yang Zhigang, Peng Taijiang, Cheng Guangming, and Wu~Boda.
\newblock Design and test of a high-performance piezoelectric micropump for
  drug delivery.
\newblock {\em Sensors and Actuators A: Physical}, 121(1):156--161, 2005.

\bibitem{cui2007study}
Qifeng Cui, Chengliang Liu, and Xuan~F Zha.
\newblock Study on a piezoelectric micropump for the controlled drug delivery
  system.
\newblock {\em Microfluidics and Nanofluidics}, 3(4):377--390, 2007.

\bibitem{lintel1988piezoelectric}
HTG~van Lintel, FCM Pol, and S~Bouwstra.
\newblock A piezoelectric micropump based on micromachining of silicon.
\newblock {\em Sensors and actuators}, 15(2):153--167, 1988.

\bibitem{sim2003phase}
Woo~Young Sim, Hyeun~Joong Yoon, Ok~Chan Jeong, and Sang~Sik Yang.
\newblock A phase-change type micropump with aluminum flap valves.
\newblock {\em Journal of Micromechanics and Microengineering}, 13(2):286,
  2003.

\bibitem{kim2005fabrication}
Ki~Hoon Kim, Hyeun~Joong Yoon, Ok~Chan Jeong, and Sang~Sik Yang.
\newblock Fabrication and test of a micro electromagnetic actuator.
\newblock {\em Sensors and Actuators A: Physical}, 117(1):8--16, 2005.

\bibitem{yun2002surface}
Kwang-Seok Yun, Il-Joo Cho, Jong-Uk Bu, Chang-Jin Kim, and Euisik Yoon.
\newblock A surface-tension driven micropump for low-voltage and low-power
  operations.
\newblock {\em Journal of microelectromechanical systems}, 11(5):454--461,
  2002.

\bibitem{richter1991micromachined}
A~Richter, A~Plettner, KA~Hofmann, and H~Sandmaier.
\newblock A micromachined electrohydrodynamic (ehd) pump.
\newblock {\em Sensors and Actuators A: Physical}, 29(2):159--168, 1991.

\bibitem{tsai2002thermal}
Jr-Hung Tsai and Liwei Lin.
\newblock A thermal-bubble-actuated micronozzle-diffuser pump.
\newblock {\em Journal of microelectromechanical systems}, 11(6):665--671,
  2002.

\bibitem{kargov2008development}
A~Kargov, T~Werner, C~Pylatiuk, and S~Schulz.
\newblock Development of a miniaturised hydraulic actuation system for
  artificial hands.
\newblock {\em Sensors and Actuators A: Physical}, 141(2):548--557, 2008.

\bibitem{yang1999design}
Xing Yang, Charles Grosjean, and Yu-Chong Tai.
\newblock Design, fabrication, and testing of micromachined silicone rubber
  membrane valves.
\newblock {\em Journal of microelectromechanical systems}, 8(4):393--402, 1999.

\bibitem{rapp1994liga}
R~Rapp, WK~Schomburg, D~Maas, J~Schulz, and W~Stark.
\newblock Liga micropump for gases and liquids.
\newblock {\em Sensors and Actuators A: Physical}, 40(1):57--61, 1994.

\bibitem{van1990thermopneumatic}
FCM Van~de Pol, HTG Van~Lintel, M~Elwenspoek, and JHJ Fluitman.
\newblock A thermopneumatic micropump based on micro-engineering techniques.
\newblock {\em Sensors and Actuators A: Physical}, 21(1):198--202, 1990.

\bibitem{zengerle1995bidirectional}
R~Zengerle, J~Ulrich, S~Kluge, M~Richter, and A~Richter.
\newblock A bidirectional silicon micropump.
\newblock {\em Sensors and Actuators A: Physical}, 50(1):81--86, 1995.

\bibitem{shen2008high}
M~Shen, C~Yamahata, and MAM Gijs.
\newblock A high-performance compact electromagnetic actuator for a pmma
  ball-valve micropump.
\newblock {\em Journal of Micromechanics and Microengineering}, 18(2):025031,
  2008.

\bibitem{yao2003porous}
Shuhuai Yao, David~E Hertzog, Shulin Zeng, James~C Mikkelsen, and Juan~G
  Santiago.
\newblock Porous glass electroosmotic pumps: design and experiments.
\newblock {\em Journal of Colloid and Interface Science}, 268(1):143--153,
  2003.

\bibitem{reichmuth2003increasing}
David~S Reichmuth, Gabriela~S Chirica, and Brian~J Kirby.
\newblock Increasing the performance of high-pressure, high-efficiency
  electrokinetic micropumps using zwitterionic solute additives.
\newblock {\em Sensors and Actuators B: Chemical}, 92(1):37--43, 2003.

\bibitem{dopper1997micro}
J~D{\"o}pper, M~Clemens, W~Ehrfeld, S~Jung, KP~Kaemper, and H~Lehr.
\newblock Micro gear pumps for dosing of viscous fluids.
\newblock {\em Journal of Micromechanics and Microengineering}, 7(3):230, 1997.

\bibitem{totten2011handbook}
George~E Totten.
\newblock {\em Handbook of hydraulic fluid technology}.
\newblock CRC Press, 2011.

\bibitem{merritt1967hydraulic}
Herbert~E Merritt.
\newblock {\em Hydraulic control systems}.
\newblock John Wiley \& Sons, 1967.

\bibitem{hooke1984end}
CJ~Hooke and E~Koc.
\newblock End plate balance in gear pumps.
\newblock {\em Proceedings of the Institution of Mechanical Engineers, Part B:
  Journal of Engineering Manufacture}, 198(1):55--60, 1984.

\bibitem{furst1998particle}
Eric~M Furst and Alice~P Gast.
\newblock Particle dynamics in magnetorheological suspensions using
  diffusing-wave spectroscopy.
\newblock {\em Physical Review E}, 58(3):3372, 1998.

\bibitem{tao1998structural}
R~Tao and Qi~Jiang.
\newblock Structural transitions of an electrorheological and
  magnetorheological fluid.
\newblock {\em Physical Review E}, 57(5):5761, 1998.

\bibitem{zhu1996role}
Yun Zhu, E~Haddadian, T~Mou, Mark Gross, and Jing Liu.
\newblock Role of nucleation in the structure evolution of a magnetorheological
  fluid.
\newblock {\em Physical Review E}, 53(2):1753, 1996.

\bibitem{jolly1999properties}
Mark~R Jolly, Jonathan~W Bender, and J~David Carlson.
\newblock Properties and applications of commercial magnetorheological fluids.
\newblock {\em Journal of intelligent material systems and structures},
  10(1):5--13, 1999.

\bibitem{goncalves2006review}
Fernando~D Goncalves, Jeong-Hoi Koo, and Mehdi Ahmadian.
\newblock A review of the state of the art in magnetorheological fluid
  technologies-part i: Mr fluid and mr fluid models.
\newblock {\em The Shock and Vibration Digest}, 38(3):203--219, 2006.

\bibitem{ginder1996rheology}
JM~Ginder, LC~Davis, and LD~Elie.
\newblock Rheology of magnetorheological fluids: models and measurements.
\newblock {\em International journal of modern physics b},
  10(23n24):3293--3303, 1996.

\bibitem{lee2000control}
Hwan-Soo Lee and Seung-Bok Choi.
\newblock Control and response characteristics of a magneto-rheological fluid
  damper for passenger vehicles.
\newblock {\em Journal of Intelligent Material Systems and Structures},
  11(1):80--87, 2000.

\bibitem{kwok2007bouc}
NM~Kwok, QP~Ha, MT~Nguyen, J~Li, and B~Samali.
\newblock Bouc--wen model parameter identification for a mr fluid damper using
  computationally efficient ga.
\newblock {\em ISA transactions}, 46(2):167--179, 2007.

\bibitem{kim1999vibration}
Kiduck Kim and Doyoung Jeon.
\newblock Vibration suppression in an mr fluid damper suspension system.
\newblock {\em Journal of Intelligent Material Systems and Structures},
  10(10):779--786, 1999.

\bibitem{sassi2005innovative}
Sadok Sassi, Khaled Cherif, Lotfi Mezghani, Marc Thomas, and Asma Kotrane.
\newblock An innovative magnetorheological damper for automotive suspension:
  from design to experimental characterization.
\newblock {\em Smart Materials and Structures}, 14(4):811, 2005.

\bibitem{yoo2002design}
Jin-Hyeong Yoo and Norman~M Wereley.
\newblock Design of a high-efficiency magnetorheological valve.
\newblock {\em Journal of Intelligent Material Systems and Structures},
  13(10):679--685, 2002.

\bibitem{guo2003finite}
NQ~Guo, H~Du, and WH~Li.
\newblock Finite element analysis and simulation evaluation of a
  magnetorheological valve.
\newblock {\em The international journal of advanced manufacturing technology},
  21(6):438--445, 2003.

\bibitem{yokota1999pressure}
Shinichi YOKOTA, Kazuhiro YOSHIDA, and Yutaka KONDOH.
\newblock A pressure control valve using mr fluid.
\newblock In {\em Proceedings of the JFPS international symposium on fluid
  power}, volume 1999, pages 377--380, 1999.

\bibitem{rowe1970measurements}
M\_ Rowe.
\newblock Measurements and computations of flow in pipe bends.
\newblock {\em Journal of Fluid Mechanics}, 43(04):771--783, 1970.

\bibitem{melle2003microstructure}
Sonia Melle, Oscar~G Calder{\'o}n, Miguel~A Rubio, and Gerald~G Fuller.
\newblock Microstructure evolution in magnetorheological suspensions governed
  by mason number.
\newblock {\em Physical Review E}, 68(4):041503, 2003.

\bibitem{du2016modified}
Di~Du, Elaa Hilou, and Sibani~Lisa Biswal.
\newblock Modified mason number for charged paramagnetic colloidal suspensions.
\newblock {\em Physical Review E}, 93(6):062603, 2016.

\bibitem{van2014microfluidics}
Albert Van~den Berg and Loes Segerink.
\newblock {\em Microfluidics for medical applications}, volume~36.
\newblock Royal Society of Chemistry, 2014.

\bibitem{becnel2014mason}
Andrew~C Becnel, Wei Hu, and Norman~M Wereley.
\newblock Mason number analysis of a magnetorheological fluid-based rotary
  energy absorber.
\newblock {\em IEEE Transactions on Magnetics}, 50(11):1--4, 2014.

\bibitem{becnel2015nondimensional}
Andrew~C Becnel, Stephen Sherman, Wei Hu, and Norman~M Wereley.
\newblock Nondimensional scaling of magnetorheological rotary shear mode
  devices using the mason number.
\newblock {\em Journal of Magnetism and Magnetic Materials}, 380:90--97, 2015.

\bibitem{sherman2015relating}
Stephen~G Sherman, Andrew~C Becnel, and Norman~M Wereley.
\newblock Relating mason number to bingham number in magnetorheological fluids.
\newblock {\em Journal of Magnetism and Magnetic Materials}, 380:98--104, 2015.

\bibitem{gjerstad2012simplified}
Kristian Gjerstad, Rune~W Time, and Knut~S Bj{\o}rkevoll.
\newblock Simplified explicit flow equations for bingham plastics in
  couette--poiseuille flow--for dynamic surge and swab modeling.
\newblock {\em Journal of Non-Newtonian Fluid Mechanics}, 175:55--63, 2012.

\bibitem{liu2010axial}
Yu-Quan Liu and Ke-Qin Zhu.
\newblock Axial couette--poiseuille flow of bingham fluids through concentric
  annuli.
\newblock {\em Journal of Non-Newtonian Fluid Mechanics}, 165(21):1494--1504,
  2010.

\bibitem{menon2004piping}
Shashi Menon.
\newblock {\em Piping calculations manual}.
\newblock McGraw Hill Professional, 2004.

\end{thebibliography}
\bibliographystyle{unsrt}
\end{document}